\documentstyle[preprint,aps]{revtex}
\tightenlines
\begin{document}

\title{Nonlinear Effects In Site Blocking 
Induced Oscillations}
\draft
\author{Dorjsuren Battogtokh$^{1,2}$ and Ravjir Davaanyam$^2$}
\address{$^1$Deparment of Biology, Virginia Polytechnik 
Institute and State University, Blacksburg, VA 24061\\
$^2$Physics and Technology Institute, 
Mongolian Academy of Sciences,Ulaanbaatar 51, Mongolia}
\date{\today}
\maketitle
\begin{abstract}
Higher order nonlinear effects of site blocking, 
which induces oscillations in the Monte Carlo 
simulation of CO oxidation, are outlined here. It is shown 
that the rate equations which include these 
effects exhibit a supercritical  Hopf bifurcation
in parameter domains where 
Monte Carlo simulations lead to oscillations.

\end{abstract}


\section{Introduction}
Surface chemical reactions play an important role in the processes
of heterogeneous catalysis, which are widely used in the chemical
industry~\cite{ASM}. Monte Carlo simulation is a powerful tool 
for a microscopic description of these reactions~\cite{Gil,David}. 
Kinetic phase transitions, oscillations and chaos, pattern 
formation,
and coupling between catalytic oscillators have been studied
by the Monte Carlo method~\cite{ZGB,Kortluke,Zhdanov,Pattern}.

The oxidation of carbon monoxide $CO$ is one
of the most extensively studied
surface catalytic reactions~\cite{Ronald,Imbihl}.
In a recent Monte Carlo study, an attenpt of a mean field
modeling was made for site 
blocking induced surface coverage oscillations~\cite{Jansen}. 
The authors of Ref.~\cite{Jansen} studied  
a mathematical model based on macroscopic rate equations for the
characterization of oscillations seen in the Monte Carlo 
simulations. However, it was
found that the mean field  model does not lead to
oscillatory dynamics in a parameter region where their
Monte Carlo simulations show oscillations.
In order to obtain oscillations, they 
numerically solved a part of their rate equations with
data from the Monte Carlo simulation for dynamics
of the free site and the inert molecule coverage. 
Such a combination of
deterministic and stochastic time evolutions seems not to be
satisfactory; it is more desirable to find a supercritical 
Hopf bifurcation in the model 
to ensure the occurrence and 
stability of  oscillations. 
The aim of this work 
is to show that inclusion of  high order nonlinear terms,
which represent interactions inherent in this  system,
allows us to find oscillations within the rate equations.
We find a Hopf bifurcation in the extended system, and 
we can determine the parameter domain for oscillatory dynamics.

\section{Site Blocking Induced Oscillations}
The model system under study is the ZGB model~\cite{ZGB},
which is extended by the absorption and desorption of an 
inert molecule:

\begin{eqnarray}
CO_{gas}+* \stackrel{w_1}{\rightarrow}CO_{ads},\\\nonumber
CO_{ads}\stackrel{w_2}{\rightarrow}CO_{gas} + *,\\\nonumber
O_2+2*\stackrel{w_4}{\rightarrow}2O_{ads},\\\nonumber
CO_{ads}+O_{ads}\stackrel{w_3}{\rightarrow}CO_2+2*,\\\nonumber
M_{gas}+*\stackrel{w_5}{\rightarrow}M_{ads},\\\nonumber
M_{ads}\stackrel{w_6}{\rightarrow}M_{gas}+*,
\end{eqnarray}
where $ads$ indicates that the molecule is absorbed on the
surface and  $gas$ indicates that the molecule is in the gas phase.
$*$ indicates a vacant site and $M$ represents the site blocking
inert molecule.

As was reported by Jansen and Nieminen~\cite{Jansen},  
one can observe sustained coverage oscillations
in the Monte Carlo simulations of Eqn. (1).
In Fig. 1 we show results of
our Monte Carlo simulation for these coverage oscillations. 
All variables and time are dimensionless in this paper.
As compared with Ref.~\cite{Jansen},
we have used a different set of parameters to show
that these oscillations can be observed in wide range of parameters.
To improve the numerical performance we have implemented
{\em lists} in our Monte Carlo simulations. Separate lists for the
locations of  free sites, $CO$, and inert molecules were
updated during the simulations. For example, for an adsorption of 
a chosen adsorbent, a site was randomly selected from the 
list of free sites; for a desorption, a site 
was randomly selected from the corresponding list of $CO$ or $M$.

The authors of Ref.~\cite{Jansen} studied  rate equations
for a macroscopic characterization of the oscillations seen in the
Monte Carlo simulation. The rate equations for the coverage of 
$CO_{ads}$ - $x$, $O_{ads}$ - $y$, and $M_{ads}$ - $z$ 
are~\cite{Jansen}:
\begin{eqnarray}
\sigma=1-x-y-z,\\\nonumber
\dot x=w_1 \sigma-w_2 x-4 w_3 x y,\\\nonumber
\dot y=4 w_4 \sigma^2-4 w_{3} x y,\\\nonumber
\dot z=w_5 \sigma -w_6 z,
\end{eqnarray}
where $\sigma$ is the free site coverage. 
It is known that Eqn. (2)
shows oscillations in a narrow parameter domain~\cite{Vigil},
however, temporal pattern of these oscillations are very different from
the oscillations seen in Monte Carlo simulations as shown in Fig. 1.
It was found that 
Eqn. (2) does not show oscillations in the
parameter domain where Monte Carlo simulations
lead to oscillations. 
To obtain oscillations, 
the authors of
Ref.~\cite{Jansen} used data for $\sigma$ and $z$ from the
Monte Carlo simulation. 
We suppose that the reason why
Eqn. (2) does not show oscillations in the 
parameter domains of intereset 
is that it does not
include important higher order nonlinear effects 
inherent in Eqn. (1), which are essential
for oscillations shown in Fig. 1.

\section{Higher order Nonlinear effects}
Monte Carlo simulations of Eqn. (1) 
suggest that  two nonlinear effects are crucial for
oscillations in this system. They destabilize  $O$ 
and $CO$ rich surfaces, correspondingly.

In Fig. 2 we show snapshots of the surface at two different time
moments. Fig. 2a corresponds to the time moment when there
are no $CO_{ads}$ molecules left on the surface. 
Note that in the original ZGB 
model, for the parameters we have chosen, the system goes to the 
state fully covered by $O_{ads}$~\cite{ZGB}. 
However, the presence 
of the inert molecules shown in Fig. 2a as the brightest spots,
prevent such a fully $O_{ads}$ covered state. Moreover, these
inert molecules allow for the presence of enough vacant
sites; these are shown
as black areas in Fig. 2. In these vacant areas, 
$CO$ molecules can be
absorbed; their coverage can grow if they
are isolated by the inert molecules from $O_{ads}$ covered
areas. By taking into account this nonlinear effect, 
which includes interaction among $CO_{ads}$, $O_{ads}$, and
$M_{ads}$, the reaction term in  Eqn. (2) can be modified
as $4 w_3 x y(1- \kappa z)$, where $\kappa$ is the blocking 
coefficient. 

Fig. 2b shows nucleation of the $O_{ads}$ covered state on the
$CO_{ads}$ rich surface. As an adsorption of $O_2$ requires 
two free sites,
nucleation of the $O_{ads}$ island may seem an unlikely
process, 
if one starts from the
fully $CO_{ads}$ covered state as an initial condition. 
However, there is an another nonlinear effect
which may lead to explosive growth of an $O_{ads}$ island. 
Suppose that the desorptions of
two $CO_{ads}$ molecules have  vacated a pair of adjacent sites. 
On these sites, $O_2$ can be absorbed, and if this happens, 
$CO_{ads}$ and $O_{ads}$ will immediately react. As
the surface is rich with $CO_{ads}$, the
reaction will free four sites. On these four sites, 
two $O_2$ molecules can be absorbed. This cascade may 
lead to explosive growth of the $O_{ads}$ island. Therefore,
 there is an another  
nonlinear effect which is particularly 
noticeable during a  nucleation of an 
$O_{ads}$ island on a $CO_{ads}$ rich surface. 
In the lowest order of nonlinearity, 
due  to the
nonlinear increase 
of available adjacent free sites through the reaction,
the growth of $O_{ads}$
can be termed as $4\omega w_3 x^2 y$, where $\omega$ is 
the coefficient for the nonlinear growth of adjacent free sites. 
We verified that $4\omega w_3 x^2 y^2$ can also serve well for this term.
 
The two nonlinear effects manifest themselves
randomly in Monte Carlo simulations. For a macrospcopic
description, we consider  them deterministic processes.
Thus, Eqn. (2) can be modified into the form:

\begin{eqnarray}
\dot x=w_1 \sigma-w_2 x-4 w_3 x y(1- \kappa z),\\\nonumber
\dot y=4 w_4 \sigma^2-4 w_3 x y(1-\omega x),\\\nonumber
\dot z=w_5 \sigma -w_6 z.
\end{eqnarray}

We want Eqn. (3) as simple as possible.
We include the reaction term $4 w_3 \kappa x y z$ 
in the first equation only because its effect
in the second equation is assumed to be small. 
This assumption stems from a 
microscopic process: a pair of  $O_{ads}$
can not be blocked by a single inert molecule. 
The reason why we include $4 w_3 \omega x^2 y$ 
in the second equation only is also taken from a 
microscopic fact: for the adsorption
of $CO$, the presence of two free adjacent 
sites is not essential. 

We note that the two effects were also described by 
Jansen and Nieminen~\cite{Jansen}; however, 
they did not include
them explicitly in the rate equations. We also note 
that Eqn. (3)
is much simpler than the rate equations in the pair
approximation of  Eqn. (1), which do not show oscillations
in the parameter range of interest~\cite{Vigil}.
Eqn. (3) is an approximate model, 
and it is desireable to
derive it systematically.

\section{A Hopf Bifurcation}

Because of its nonlinear terms, 
analytic treatment of Eqn. (3) is difficult. 
Nevertheless, it can be studied
numerically. For example, one may use 
the software "XPPAUT"  by B. Ermountrout~\cite{Brad}. 

We found that for the occurrence of a Hopf bifurcation
it is not necessary that both $\kappa$ and $\omega$
were nonzero in Eqn. (3). However, we found that parameter
space for oscillations are larger if both 
$\kappa$ and $\omega$ are nonzero.

\subsection{case $\kappa \neq 0$ and $\omega \neq 0$}
In the present case, it is difficult to find
analytic expressions for the steady states of Eqn. (3).
In Fig. 3 we show a numerical integration of Eqn. (4).
Here, in contrast to  its dynamics shown in
Fig. 1, the inert molecule concentration remains high.
We suppose that the main reason for this is that
$w_3$ is finite in Fig. 3, but $w_3=\infty$ in Fig. 1.
Also, a better choice of the rate constants can be made
in order to compare  dynamics in the rate equations and 
Monte Carlo simulations~\cite{Gil}.

A Hopf bifurcation in this system is shown in Fig. 4.
Thus, the nonlinear effects we outlined in the
previous section may indeed lead to stable oscillations. 
A two parameter
bifurcation diagram is shown in Fig. 5.
We found that with the increase of $w_3$,
the parameter region for a Hopf bifurcation 
widens. We found  that such an increase 
is  accompanied by the emergence of saddle nodes. 

\subsection{case $\kappa \neq 0$ and $\omega = 0$}
In the present case, the steady states are given by,
\begin{eqnarray} 
x_0={w_2}^{-1} (w_1 z_0 -4 w_4 {z_0}^2+4 \kappa w_4 {z_0}^3),\\\nonumber
y_0=1-x_0-2 z_0,
\end{eqnarray}
where $z_0$ is a solution to
\begin{eqnarray}
w_2^2 w_4 z_0+w_3 (w_1+4 w_4 z_0 (\kappa z_0-1)\\\nonumber
(w_2(2 z_0-1)+z_0(w_1+4 w_4 z_0(\kappa z_0-1)))=0.
\end{eqnarray}

Once again we were able to analyze 
stability of $x_0$, $y_0$ and $z_0$
numerically only.
Oscillations  for the present case are shown in Fig. 6.
We note that in the present case, except the Hopf bifurcation, 
saddle node bifurcations occur in the system.

\subsection{case $\kappa = 0$ and $\omega \neq 0$}
As in the previous case, the steady states of Eqn. (3)
can be expressed through a polynomial. Again, we
were able to solve the polynomial numerically only.  
As a result, a linear stability analysis of these steady states
has turned out to be difficult. Numerically detected 
oscillations for the present case are shown in Fig. 7.
Note the exceptions  of $w_3$'s value and $\omega \neq 0$;
these are the reaction rates used in Ref.~\cite{Jansen}
for Monte Carlo simulations.
We note that for the parameters used in Fig. 6 - 7, a parameter 
domain  for the Hopf bifurcation 
was smaller in the present case then it was in the previous case.

\section{Discussions}
A microscopic study of a reaction system based 
on Monte Carlo simulations can reveal the molecular
mechanisms of catalytic processes. As an example,
in this work, such study has allowed us 
to outline the essential nonlinear processes
in site blocking caused oscillations. However,
fora  more  complete study of a reaction system,
the role of diffusion should be taken into account
~\cite{Syn}. In that sense,
Eqn. (4) can be a useful starting point 
for a reaction diffusion study of site blocking
induced oscillations. The preliminary results of a
reaction diffusion system in one dimension,
with a diffusion of $x$, $y$, and $z$, show
synchronous oscillations, stable long lived islands
and diffusion induced chemical turbulence~\cite{Sensei}.
More detailed results on this, and its comparison
to the Monte Carlo simulations, which include diffusion 
of adsorbents, will be reported elsewhere~\cite{Bat}.

Finally, we suppose that 
the nonlinear effects and stable oscillations
due to site blocking can be detected in experimental 
studies on $CO$ oxidation.


%
%
\begin{figure}
\caption{
Coverage oscillations on the square lattice. 
Open circles: $\bar \theta_{CO_{ads}}$. 
Long dashed: $\bar \theta_{O_{ads}}$, Solid lines: 
$\bar \theta_{M_{ads}}$. Rate constants:
$w_1=0.44$, $w_2=0.005$, $w_3=\infty$,
$w_4=0.56$, $w_5=0.004$, $w_6=0.004$. Time is scaled
to the total number of sites 
$S=N^2$, $N=256$.
$\bar \theta=\frac {1}{S} \sum \theta_i, 
i=CO_{ads}, O_{ads}, M_{ads}$.}
\label{1}
\end{figure}

\begin{figure}
\caption{Snapshots of the surface at two different 
moments. a.) A phase without $CO_{ads}$ coverage.
b.) Emergence
of an $O_{ads}$ island on the $CO_{ads}$ dominant surface. 
Parameters are the same as in Fig. 1.}
\end{figure}

\begin{figure}
\caption{Coverage oscillations in Eqn. (3). 
Open circles: $x$. Long dashed line: $y$. 
Solid line: $z$. 
Parameters are the same as in Fig. 1 except $w_3=5$, 
$\omega=1$, $\kappa=2$.}
\end{figure}

\begin{figure}
\caption{A Hopf bifurcation diagram. Solid line represents a 
stable steady state. Circles represent the Hopf 
bifurcation branch: filled circles show the supercritical
Hopf bifurcation, open circles show unstable oscillations.
Dashed and thin lines show an unstable
steady state. Parameters are the same as in Fig. 3.}
\end{figure}

\begin{figure}
\caption{A two parameter bifurcation diagram. 
Oscillations occur in the area marked by {\em osc}.  Other parameters 
are the same as in Fig. 3.}
\end{figure}

\begin{figure}
\caption{
Oscillations in the case $\omega =0$. 
Open circles: $x$. Long dashed line: $y$. 
Solid line: $z$. 
Parameters are: $w_1=0.2$, $w_2=0.01$, $w_3=5$,
$w_4=1$, $w_5=w_6=0.01$ and  $\kappa=5.5$}
\end{figure}

\begin{figure}
\caption{
Oscillations in the case  $\kappa =0$.
Open circles: $x$. Long dashed line: $y$. 
Solid line: $z$. 
Parameters are: $w_1=1$, $w_2=0.001$, $w_3=5$,
$w_4=0.52$, $w_5=w_6=0.0003$, and $\omega=1$.}
\end{figure}

%
%

\newpage
\begin{picture}(3.375,4.5)
\includegraphics{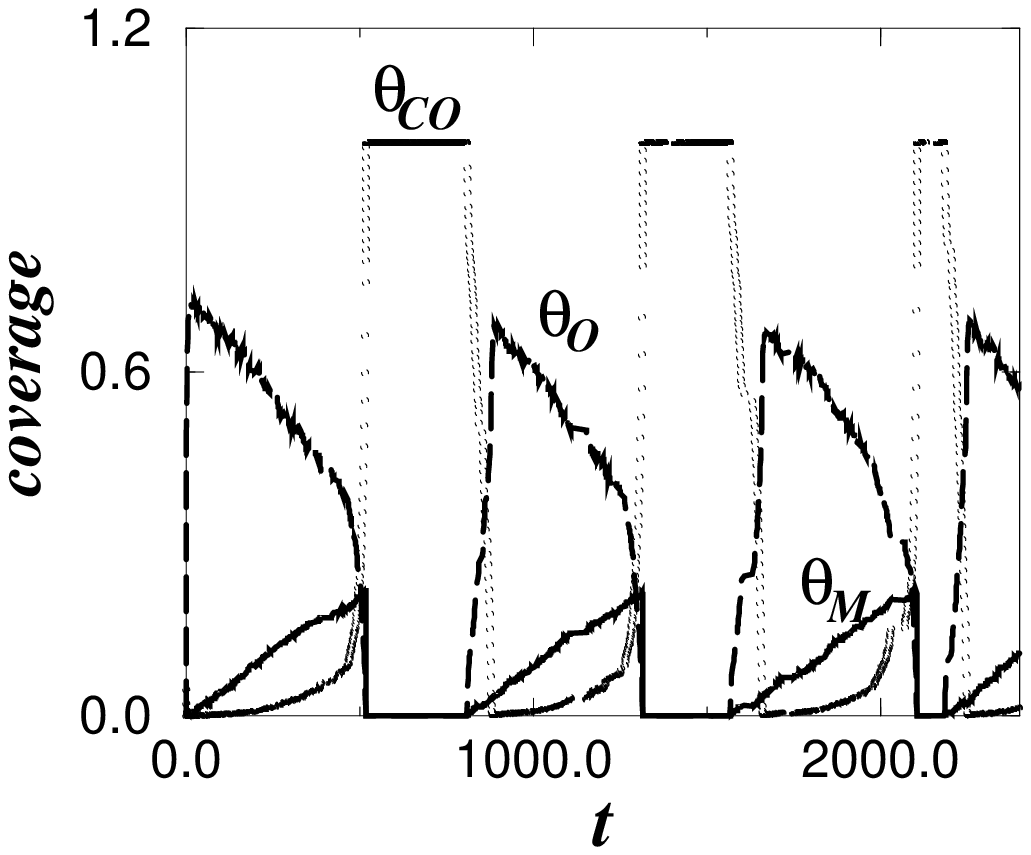}
\end{picture}

\vspace{20cm}
\centerline{\LARGE {\bf Figure 1}}

\newpage
\begin{picture}(3.375,4.5)
\includegraphics{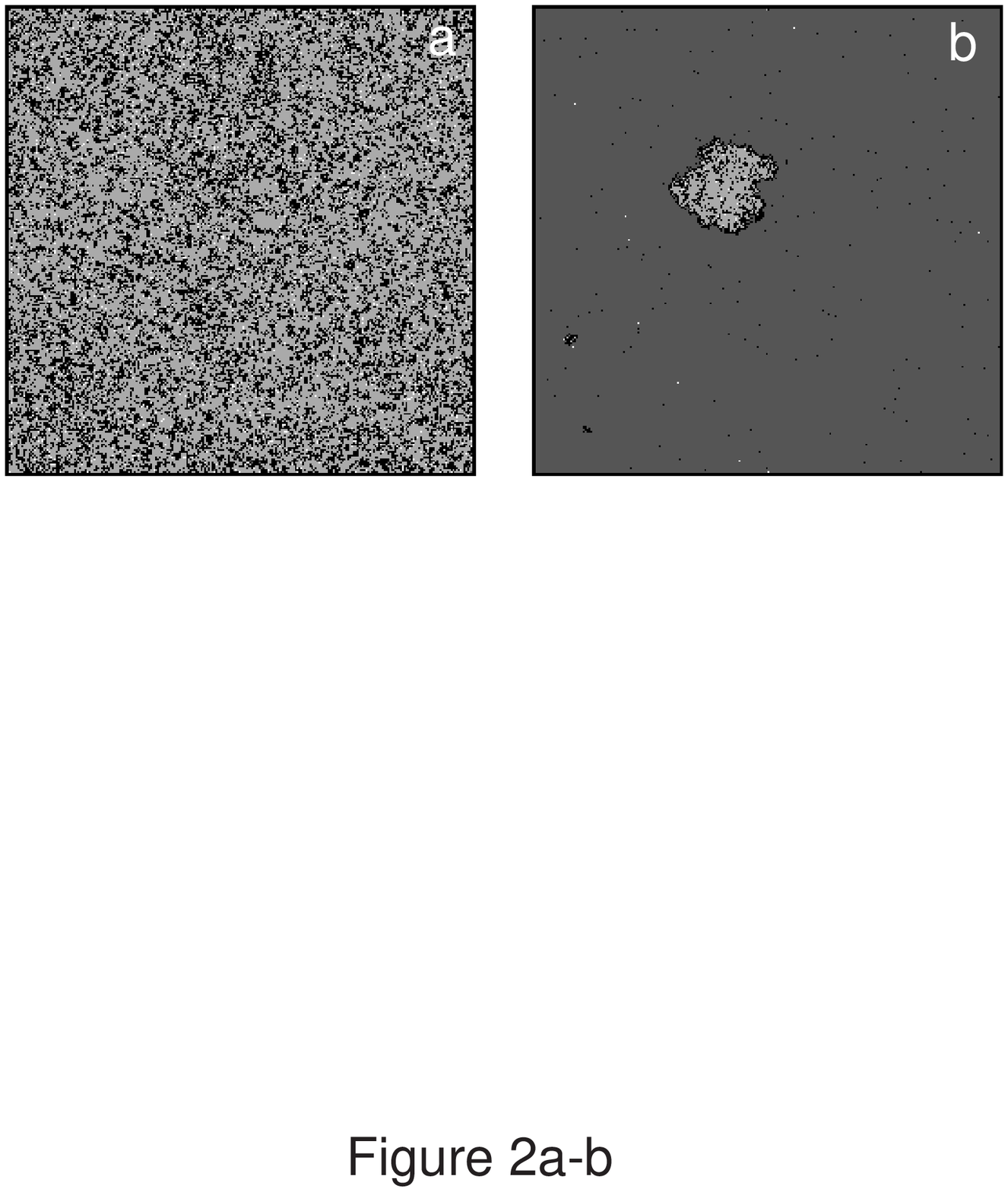}
\end{picture}

\newpage
\begin{picture}(3.375,4.5)
\includegraphics{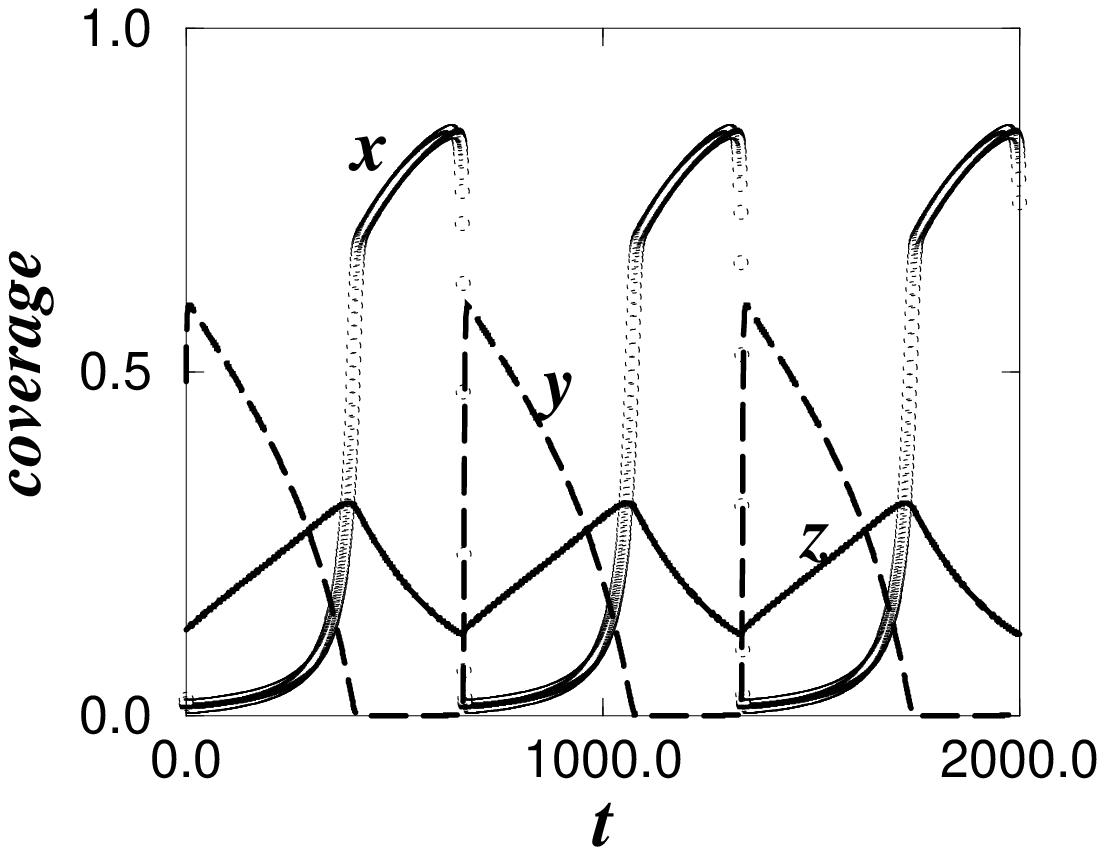}
\end{picture}

\vspace{20cm}
\centerline{\LARGE {\bf Figure 3}}

\newpage

\begin{picture}(3.375,4.5)
\includegraphics{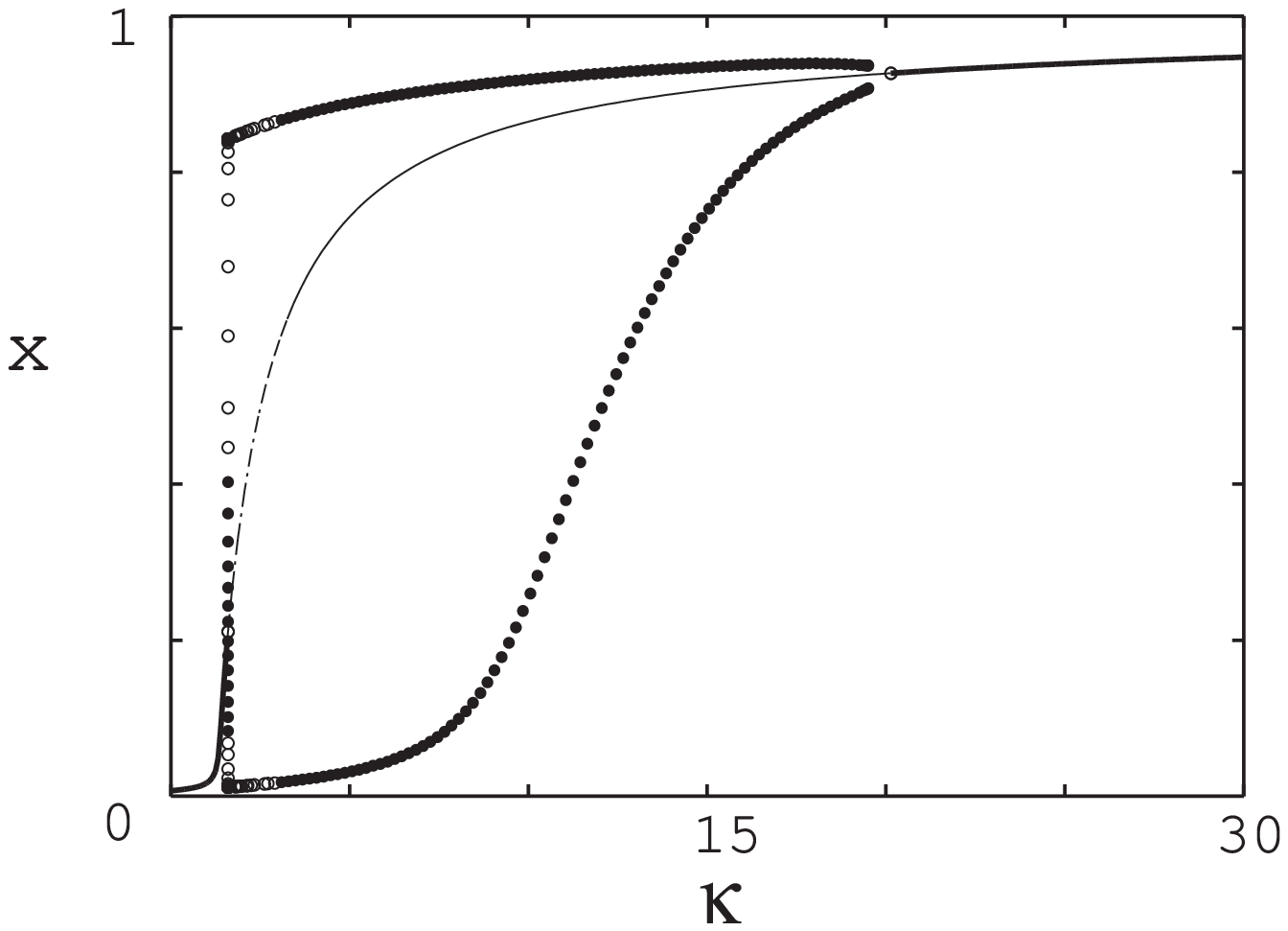}
\end{picture}

\vspace{20cm}
\centerline{\LARGE {\bf Figure 4}}

\newpage

\begin{picture}(3.375,4.5)
\includegraphics{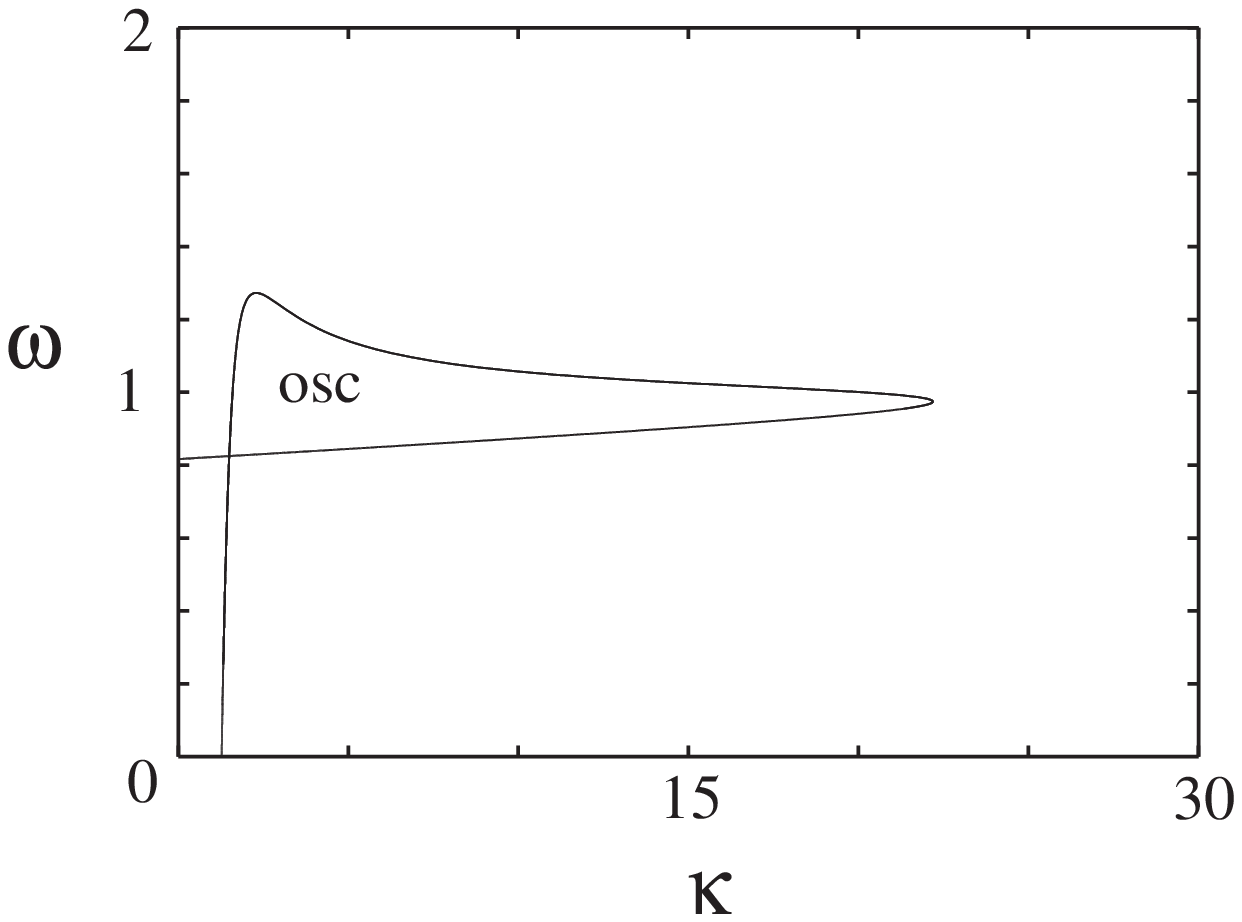}
\end{picture}

\vspace{20cm}
\centerline{\LARGE {\bf Figure 5}}

\newpage
\begin{picture}(3.375,4.5)
\includegraphics{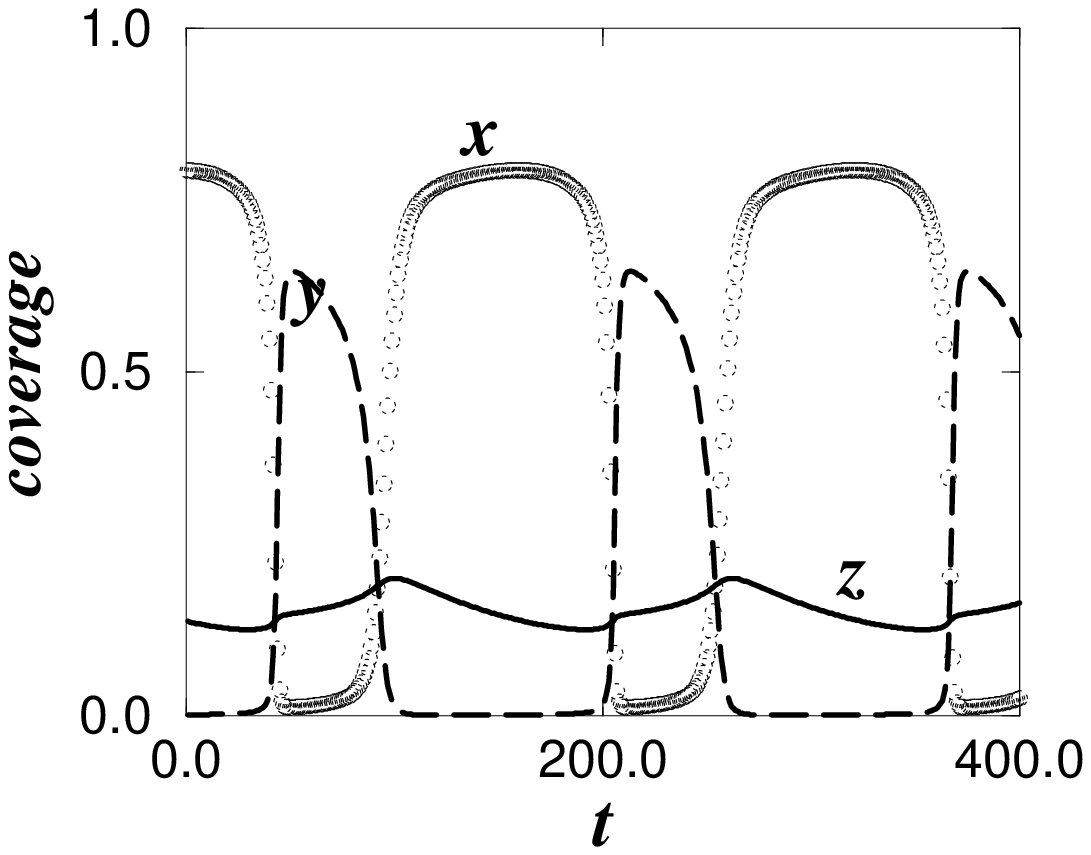}
\end{picture}

\vspace{20cm}
\centerline{\LARGE {\bf Figure 6}}

\newpage
\begin{picture}(3.375,4.5)
\includegraphics{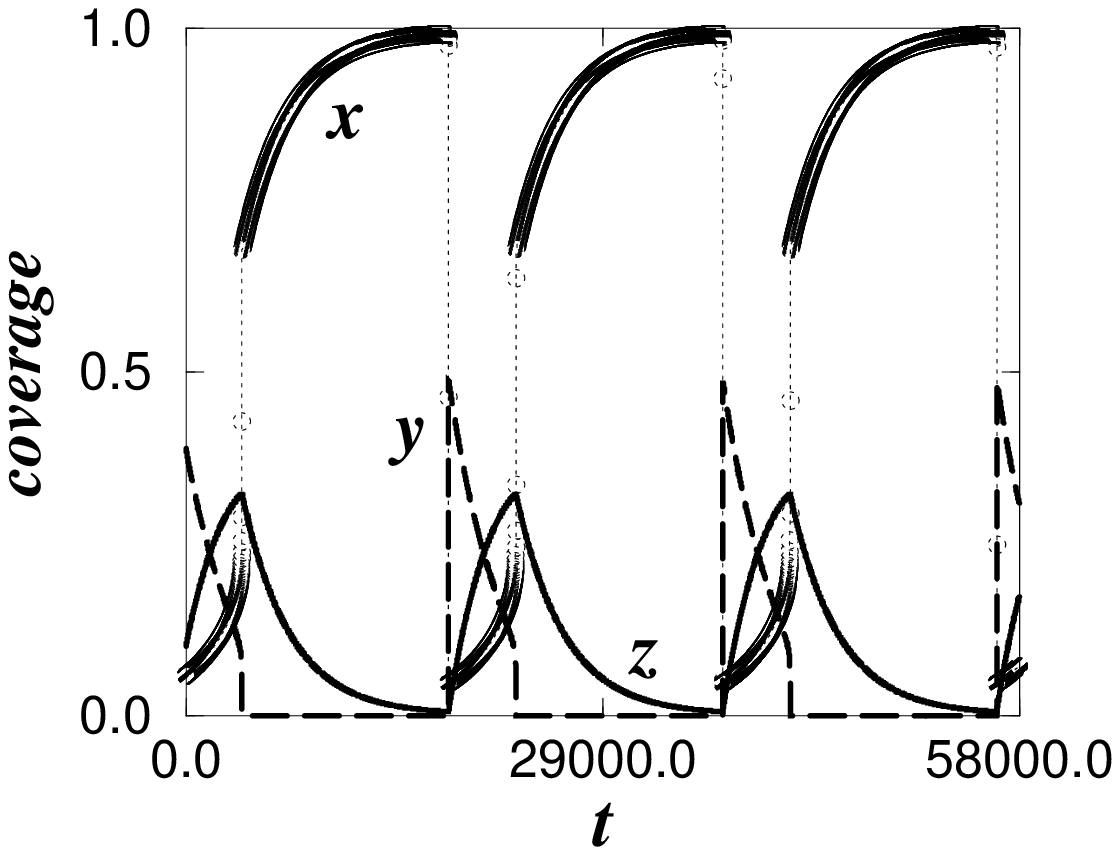}
\end{picture}

\vspace{20cm}
\centerline{\LARGE {\bf Figure 7}}

\end{document}